# The VCSEL-based Array Optical Transmitter (ATx) Development Towards 120-Gbps Link for Collider Detector: Development Update

Di Guo[a,b], Chonghan Liu[b], Jinghong Chen[c,d], John Chramowicz[e], Datao Gong[b], Suen Hou[f], Deping Huang[c], Ge Jin[a], Xiaoting Li[b,g], Tiankuan Liu[b], Alan Prosser[e], Ping-Kun Teng[f], Jingbo Ye[b], Yongzhao Zhou[a], Yang You[c], Annie C. Xiang[b] and Hao Liang[a,*]

[a] *State Key Laboratory of Particle Detection and Electronics,*
  *University of Science and Technology of China,*
  *Hefei Anhui 230026, China*

[b] *Department of Physics, Southern Methodist University,*
  *Dallas, TX 75275, USA*

[c] *Department of Electrical Engineering, Southern Methodist University,*
  *Dallas, TX 75275, USA*

[d] *Department of Electrical Engineering, University of Houston,*
  *Houston, TX 77004, USA*

[e] *Real-Time Systems Engineering Department, Fermi National Laboratory,*
  *Batavia, IL 60510, USA*

[f] *Institute of Physics, Academia Sinica,*
  *Nangang 11529, Taipei, Taiwan*

[g] *Department of Physics, Central China Normal University,*
  *Wuhan, Hubei 430079, P.R. China*
  *E-mail:* simonlh@ustc.edu.cn

ABSTRACT: A compact radiation-tolerant array optical transmitter module (ATx) is developed to provide data transmission up to 10Gbps per channel with 12 parallel channels for collider detector applications. The ATx integrates a Vertical Cavity Surface-Emitting Laser (VCSEL) array and driver circuitry for electrical to optical conversion, an edge warp substrate for the electrical interface and a micro-lens array for the optical interface. This paper reports the continuing development of the ATx custom package. A simple, high-accuracy and reliable active-alignment method for the optical coupling is introduced. The radiation-resistance of the optoelectronic components is evaluated and the inclusion of a custom-designed array driver is discussed.

KEYWORDS: Optical detector readout concepts; Radiation-hard electronics; Front-end electronics for detector readout; Lasers

---

[*] Corresponding authors.

# Contents

## 1. Introduction

The data rates needed for collider detector readout upgrades have increased rapidly in recent years [1]. As most detectors have more data to transmit from the front-end to the back-end, there is a greater need for high density transmitters. For this we proposed the development of the ATx, an array optical transmitter, to provide data rates up to 10 Gbps per channel with 12 parallel channels. Single channel receivers, such as the VTRx [2], are available for data transmission in the other direction. VCSEL array is the most commonly used light source for parallel transmission. In order to use the surface emitter, there needs to be a 90 degree turn inside the transmitter package so that the emitted light can be parallel to the package base for fiber ribbon alignment. The ATx integrates a VCSEL array, an array driving ASIC, a carrier substrate and array optics components in a custom package. The ATx supports multi-mode operation at 850 nm wavelength over 100 meters of OM3 or OM4 fibers and works with commercial parallel receivers at the back-end. Prior work has validated several ATx custom packages with 10 Gbps per channel transmission [3].

    One critical step of packaging the array transmitter module is the optical alignment process. It is a sensitive parameter of the link performance, directly affecting the coupling efficiency. As the maturity of photonic device packaging is not yet comparable to that of microelectronics packaging, it is important to understand the technical and economic challenges of these processes. We have previously partnered with an assembly supplier to evaluate a visually controlled passive alignment process. With the potential for automated volume production, the low coupling penalty yield is yet to be improved. Alternatively, we have explored the more flexible active-alignment process. The coupling efficiency can be monitored and controlled throughout the process and the time to achieve optimization is quite affordable for production in small quantities.

    The ATx needs to be radiation resistant for collider detector front-end applications. The in-house customization capability is thus necessary as each of the constituent components need to

be radiation qualified. VCSELs have demonstrated tolerance to Non-Ionizing Energy Loss (NIEL) up to $10^{15}$ cm$^{-2}$ 1-MeV neutrons equivalent, with increased threshold, decreased slope efficiency, but comparable frequency response [4]. The optical assembly needs be tested with the inclusion of the VCSEL, whose degradation is known. The driver ASIC however needs to be tested in the ATx custom package optically. The ATx module can be customized to accommodate different array drivers. A commercial 12-channel, 10 Gbps/ch array driver was used last year, and now it is replaced by a custom VCSEL array driving ASIC (LOCld4) [5]. The LOCld4 is a four channel radiation-tolerant VCSEL array driver fabricated in a commercial 0.25-µm Silicon-on-Sapphire (SoS) CMOS process [6] with each channel operating at 8 Gbps.

The structure of the ATx is introduced in terms of the optical interface and the electrical interface in Section 2. The procedures of the in-house active-alignment are described in Section 3. In Section 4, we discuss the ATx test results including optical coupling, the X-ray irradiation and the custom ASIC integration. The conclusions are presented in Section 5.

## 2. Optical and electrical interfaces of the ATx

The ATx is a 12 channel transmitter employing commercially available optical coupling elements in a custom package. The bare dies of the VCSEL array and driving ASIC are attached and wire bonded to the substrate, which has an edgewrap structure around the periphery as the electrical interface. The array optical components are assembled above the two dies, working as the optical interface to couple optical outputs of 12 channels into the fiber ribbon. Figure 1(a) is a photo showing a fully assembled ATx module.

The array optical components consist of two parts: the Module Optical Interface (MOI) and the Prizm fiber. PRIZM and LightTurn are trademarks of US Conec Ltd. The MOI needs to be attached on the substrate over the VCSEL array die, and has 12 micro-lens in the center to collimate the 12 light outputs emitted from the VCSEL array. The Prizm fiber is a 12-channel fiber ribbon with one end terminated with a Prizm Connector, and the other end terminated with a Multiple Termination Push-on (MTP) connector [7]. The MOI and the Prizm fiber are shown in Figure 1(b). Light outputs of 12 channels collimated by the MOI will be distributed into 12 total internal reflection (TIR) microprisms in the Prizm Connector, deflected 90 degrees by TIR, and coupled into the fiber ribbon, as shown in Figure 1(c). The Prizm Connector can be clipped onto the MOI, and alignment between the MOI and the Prizm Connector is guaranteed by two precise guide pins on the Prizm Connector and guide pin holes in the MOI. Therefore, aligning the MOI with the VCSEL array becomes critical.

The VCSEL array die, the driving ASIC die and the MOI are attached on the substrate which has an "edgewrap" structure around the four sides to constitute the electrical interface. A 3-D model of the substrate is shown in Figure 2(a). The ATx module needs to be soldered onto the mother board through the "edgewrap". Another version of the ATx module with a pluggable connector is also designed [3]. The substrate is 1.9 cm × 2.2 cm × 0.9 mm in size, and the lateral dimensions of a complete ATx module with MOI and Prizm connector are shown in Figure 2(b). The first ATx module prototype integrated with a commercial VCSEL array driver has been developed and tested last year. The 10 Gbps optical eye diagram passed the eye mask, and a bit-error rate (BER) less than $10^{-12}$ transmission was achieved at 10 Gbps/channel, which proved function of the electrical interface and the optical interface at a speed up to 10 Gbps/channel.

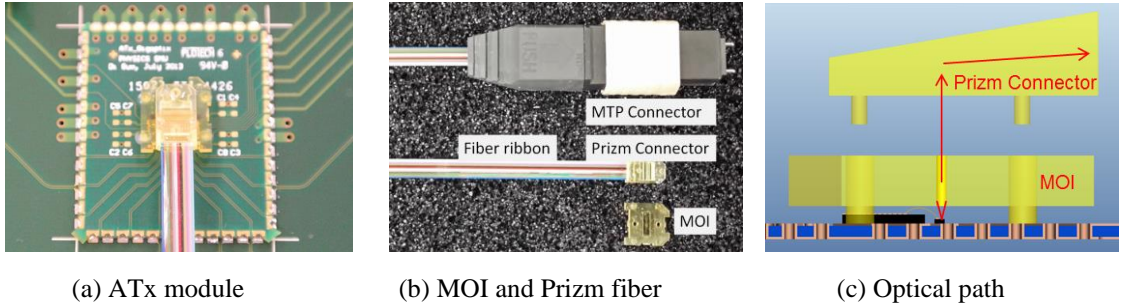

(a) ATx module      (b) MOI and Prizm fiber      (c) Optical path

**Figure 1**. Optical path and photos of MOI and Prizm connector

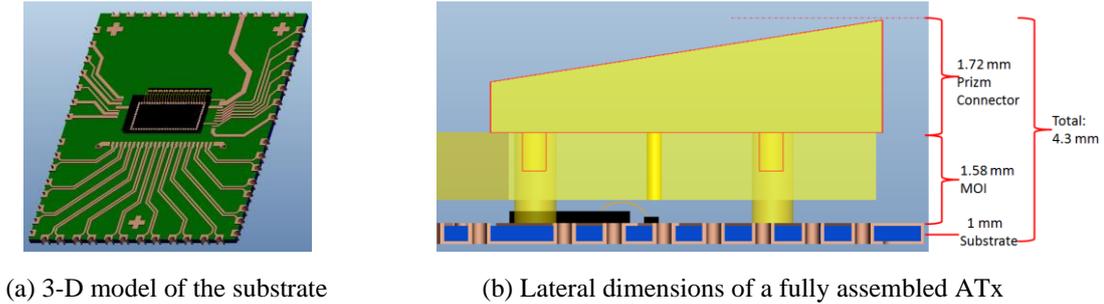

(a) 3-D model of the substrate      (b) Lateral dimensions of a fully assembled ATx

**Figure 2**. ATx substrate and dimensions

## 3. Active-alignment Method

The ATx module assembly process, including die attachment, wire bonding, MOI assembly (alignment and attachment processes), and reflow soldering, consists of crucial aspects of the ATx development. The MOI alignment and attachment process, which directly affects the final coupling efficiency, is the most challenging and critical step. The diameter of each light emitting aperture in the VCSEL array is smaller than 20 μm, and that of each micro lens in the MOI is no larger than 50 μm. To get an effective coupling efficiency of all 12 channels, the alignment tolerances on the positioning of the lens array (MOI) relative to the aperture array (VCSEL array) need to be around 10 μm. Considering the rigorous tolerance requirement, it is almost impossible to utilize any auxiliary mark on the substrate in this case. The border of the MOI is also not accurate enough for any fit-in design. Besides, the existence of the lens in the MOI makes it impossible to see through them to align with the light apertures, which aggravates the difficulty of the alignment process.

    The commercial solutions for this alignment issue include two approaches. One is to use the microscope reticles to identify positions of the lens array and the aperture array respectively, and to move the MOI to make two reticles coincident to avoid the out-of-focus issue. Last year's ATx modules were aligned in such a way. Out of all the four modules assembled last year, one module had inaccurate alignment with poor coupling efficiency, and one module was found completely misaligned. Another commercial approach is to utilize a suite of costly split screen. First, the pick and place machines drive the MOI to the pool to load the epoxy. During the alignment, the lens array (MOI) and the aperture array (VCSEL array) are viewed respectively from the top and bottom cameras. Once aligned, the MOI continues to be driven in the z-direction to the substrate and left there for the epoxy curing. The system to implement this solution costs several hundred thousand dollars which is too expensive for research and development. Despite the high cost, when evaluating this kind of alignment solution, we still

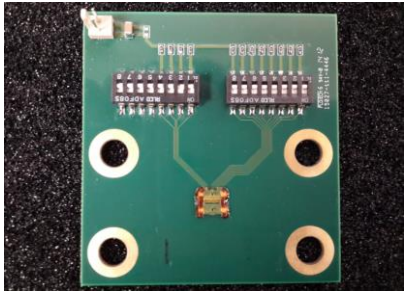 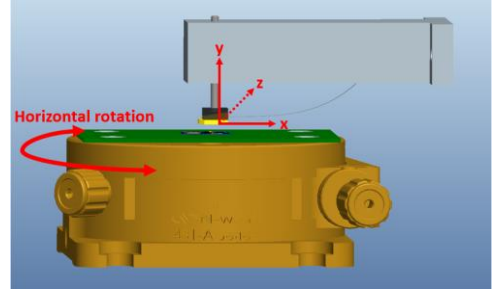

(a) Active-alignment evaluation board  (b) Movement method of the MOI and the substrate

**Figure 3**. Evaluation board and movement method

found failed examples. The evaluation result will be shown in the next section. According to the feedback and our experience, the failure mainly happened in the epoxy curing process. The potential tiny shift of the MOI during the curing process is enough to violate the tolerance requirement, and makes the assembly more unpredictable even after the MOI is aligned. Moreover, different amounts, positions and materials of the epoxy result in widely varying alignments, which compounds the problem.

In order to avoid these unstable factors and meet the strong need for a low-cost system for research and development, an active-alignment method has been developed. First, an Active-alignment Evaluation Board, shown in Figure 3(a), was designed. The 12-channel VCSEL array was attached on the board and DC driven without drivers. Each channel can be turned on or off independently. The board was fixed onto a precise rotation stage so that the VCSEL array (aperture array) could be freely rotated horizontally. A mechanical arm and fixture was created to hold the Prizm Connector in the nano-positioner. The MOI (lens array) was clipped onto the Prizm Connector so that the lens array could freely move in x, y, and z directions through the nano-positioner. Figure 3(b) is a 3-D model showing the movements (the nano-positioner is not included in the picture). During the alignment process, the VCSEL array was turned on. The alignment of the MOI over the VCSEL array was optimized based on the optical power. For a 12 channel or 4 channel VCSEL array, only two end channels were read out to guide the alignment process. The VCSEL array was accurate enough to guarantee that all the channels were aligned once the two end channels were aligned.

Figure 4(a) depicts the alignment process guided by the optical output of two end channels (channel A and channel B). It is highly procedural after one end channel (channel A) is roughly aligned: the target of a rotation step is to increase the power of channel B and to decrease that of the channel A, while the target of a translation step is to increase the power of channel A to make up for the rotation step. Two steps continue until both end channels get desirable outputs. The automatic alignment under computer control is under development. However, it takes no longer than 10 minutes per module even if the manual stages are used.

Once alignment is finished, the MOI will be lowered down to the substrate and held by the fixture until it is epoxied down to the substrate. Fixing the MOI when applying and curing the epoxy greatly improves the consistency of the final coupling efficiency between different channels and modules. Moreover, it is possible to further adjust, if needed, the MOI after applying the epoxy according to the output power. Figure 4(b) shows the setup of the active-alignment method. When aligning the MOI with the VCSEL array in a real ATx module, a carrier board fixed on the rotation stage was used to hold and DC drive the ATx module. Now the whole MOI assembly process can be finished within 15 minutes with low-cost setup in the lab. The coupling insertion loss of each channel, the channel-to-channel variation and the light crosstalk within the adjacent channel can be stably controlled to be less than 3 dB, under 1 dB, and below -50 dBm, respectively.

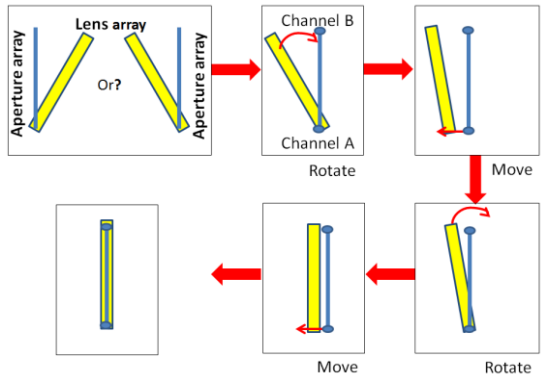 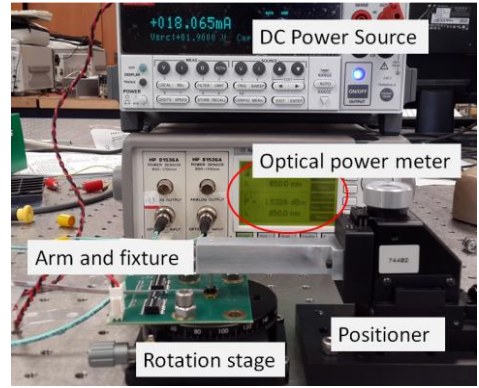

(a) Alignment steps　　　　　　　　(b) Setup of the active-alignment

**Figure 4**. Active-alignment method

In general, the active-alignment method avoids the out-of-focus issue and the costly camera system, and the coupling efficiency can be monitored all the way from the beginning of the alignment to the final attachment. The stability of the epoxy curing process is enhanced. The yield is tremendously improved by using the active-alignment method as compared to our experience of using those passive techniques.

## 4. Testing Results

### 4.1 Coupling efficiency test of the active-alignment method

The active-alignment evaluation board, on which the VCSEL array was DC driven, was used for the coupling efficiency test. The VCSEL array used in the test was ULM850-10-TT-N0112U from ULM. Two test boards were assembled with active-alignment, and all channels were DC driven at 9.15 mA, corresponding to a 4.77 dBm output according to the VCSEL array datasheet. Another two boards were assembled with passive alignment, operated in the same condition, and used the same fiber ribbon and MTP-to-LC fan out fiber. All outputs of the four boards were recorded and shown in Figure 5. The optical outputs of all 12 channels of the two boards assembled with active-alignment were above 1.8 dBm. The insertion loss of every channel of the actively aligned modules was less than 3 dB, and the channel-to-channel variation was less than 1 dB. Both are much better than those of the passively aligned modules. The insertion loss performance was also improved when compared to that of the best passively aligned ATx module from last year, which was 3 ~ 3.7 dB.

### 4.2 X-ray irradiation tests of optical components

The radiation tolerance of the optical components (the MOI and the Prizm Connector) was preliminarily tested last year due to the limited quantity of modules with successful alignment. Due to the mechanical mismatch between the MOI and Prizm Connector, up to 2 dB of post radiation penalty was reported. The MOI and the Prizm Connector were connected through two guide pins and two guide pin holes. The fiber ribbon, which connected to the Prizm Connector, applied unexpected force to the connector before the fiber ribbon was fixed. This force was big

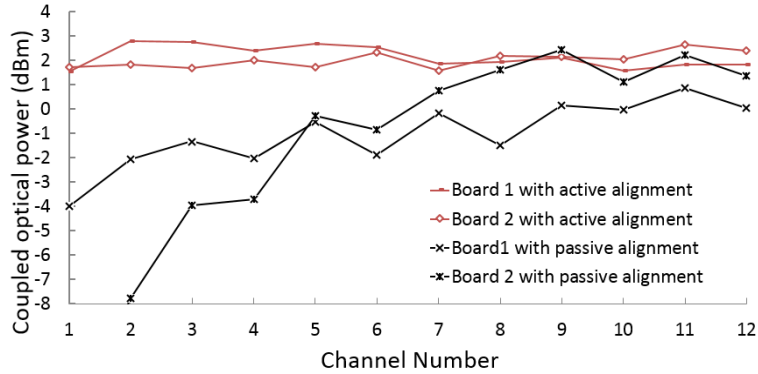

**Figure 5**. Comparison between the active-alignment and the passive alignment

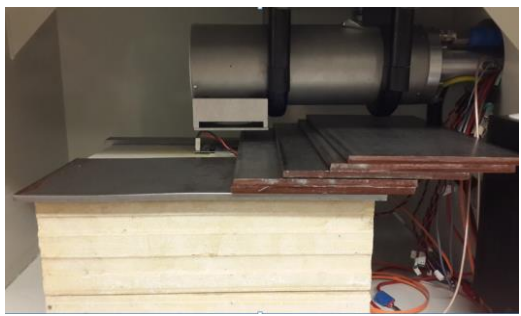

(a) Photo of the X-ray irradiation test

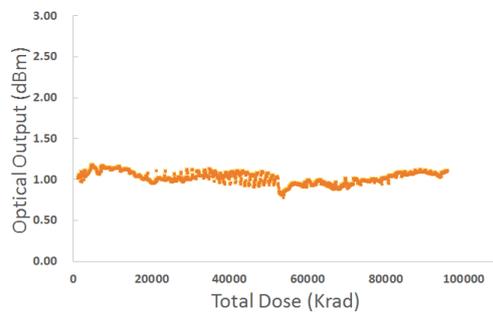

(b) Optical output versus total ionizing dose

**Figure 6**. Irradiation test setup and the results

enough to change the mechanical connection between the connector and the MOI in such a way that the coupling efficiency fluctuated in a range of 3 dB. In last year's test, the fiber ribbon was not fixed and the optical outputs before and after irradiation were recorded separately. A more complete and persuasive test was conducted this year. The VCSEL array was DC driven during the test. The fiber ribbon was fixed to ensure the stable and consistent connection between the MOI and the Prizm Connector. The optical output of one channel was captured and recorded every thirty seconds throughout the whole irradiation test. Figure 6(a) is a photo showing the test setup this year. The MOI and the Prizm connector were irradiated at a dose rate of 6668 Krad($SiO_2$)/hr for fifteen hours. The optical ouptput is shown in the Figure 6(b). No radiation-induced attenuation was observed with the total dose of up to 96 Mrad($SiO_2$).

### 4.3 Integrating the ATx with the custom driving ASIC

As the ATx is a custom assembled package, different array drivers can be integrated. A commercial 12-channel, 10 Gbps/channel array driver was used last year, and it was replaced by a custom VCSEL array driving ASIC (LOCld4). LOCld4 is a four channel radiation-tolerant VCSEL array driver fabricated in a commercial 0.25-μm Silicon-on-Sapphire (SoS) CMOS technology with each channel operating at 8 Gbps [5]. The LOCld4 adopts the open-drain output structure for array matching. The outputs share the common power source, and drive the cathodes of the VCSEL array. Nowadays, most VCSEL arrays share the common substrates at the cathode, and channels are not isolated within the array. Another version of the LOCld4 with anode-driven outputs has been designed and will be submitted in December 2014. A preliminary test was conducted for the current LOCld4 driving one channel of a common-cathode VCSEL array. The block diagram and a picture of the test setup are shown in Figure 7. The optical eye diagram was captured at 8 Gbps and passed the eye mask test, as shown in Figure 8.

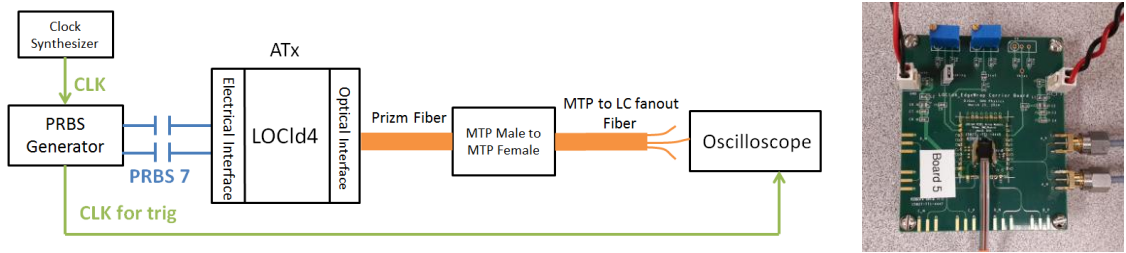

**Figure 7**. Block diagram and picture of the optical eye diagram test

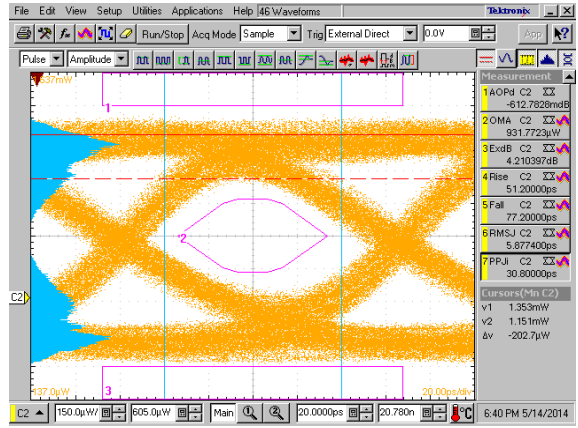

**Figure 8**. Optical eye diagram of the ATx with LOCld4 at 8 Gbps

## 5. Conclusion

This paper reports the continuing development of the ATx module for multi-channel optical transmission from detector front-end. An active-alignment method was developed in-house with which less than -3 dB insertion loss and less than 1dB channel variation can be achieved reliably. The radiation-resistance of the optoelectronic components was evaluated by x-ray irradiation up to 96 Mrad ($SiO_2$), where no coupling degradation was observed. The ATx assembly integrated with a custom VCSEL array driving ASIC (LOCld4) operating at 8 Gbps/ch was demonstrated. The complete module irradiation and full channel link performance will be evaluated.

## Acknowledgments

This work is supported by the US Department of Energy Collider Detector Research and Development (CDRD) data link program. The authors also would like to thank Jee Libres and Alvin Goats of VLISP, Michael Wiesner of ULM and Alan Ugolini of US Conec for informative discussions.